\documentclass[pra,twocolumn]{revtex4}
\usepackage[latin1]{inputenc}
\usepackage{amsfonts}
\usepackage{amssymb}
\usepackage{amstext}
\usepackage{graphicx}
\usepackage{wasysym}

\begin{document}
\title{Second Josephson oscillations}

\author{M.\,P.\,Strzys and J.\,R.\,Anglin}
\address{OPTIMAS Research Center and Fachbereich Physik, Technische Universit\"at Kaiserslautern, D--67653 Kaiserslautern, Germany}

\begin{abstract}
As a model for mesoscopic quantum systems in thermal contact, we consider a four-mode Bose-Hubbard model with two greatly differing tunneling rates.  By a series of Holstein-Primakoff transformations we show that the low-frequency dynamics of this system consists in general of two slow Josephson oscillations, rather than the single slow mode predicted by linear Bogoliubov theory.  We identify the second slow Josephson oscillation as a heat exchange mode analogous to second sound.
\end{abstract}
\maketitle

\section{Introduction}
Experiments on mesoscopic quantum systems, and sophisticated numerical techniques for quantum many body theory, promise progress on deep questions about the relationship between microphysics and thermodynamics \cite{Deut91, Sred94, Kino06, Rigo08, Ecks09, Rigo09a}. We still require analytical theory, however, to clarify the questions. For example, it should be well known that heat consists microscopically of energy held by degrees of freedom whose evolution is too rapid to perceive or control on macroscopic time scales \cite{Land80}.  This simple formulation is nonetheless surprising to many physicists, because heat is rarely discussed explicitly except in terms of ensembles. Since the ensembles themselves are justified fundamentally as proxies for time averaging over rapid microscopic evolution, the role of time scale separation in defining heat is well established as an implicit assumption; but it is underdeveloped as an explicit theory.  Investigating the mesoscopic onset of thermodynamics will require fully developing this theory, because the applicability of ensembles is not obvious in small systems.

Microscopic energy that remains adiabatically frozen in fast evolution may of course simply be ignored on longer time scales.  What the First Law of Thermodynamics means from a microscopic point of view, however, is that energy can be exchanged slowly between slow and fast degrees of freedom: heat and work may transform into each other. To investigate the mesoscopic onset of thermodynamics, therefore, a simple dynamical model that provides a minimal representation of heat exchange offers a useful paradigm.  

In this work we present a simple but non-trivial model for two quantum subsystems in thermal contact, and show how heat exchange between them may be described in familiar quantum mechanical terms.  Each subsystem is a pair of bosonic modes coupled so as to represent an idealized Josephson junction -- a two-mode Bose-Hubbard system. We specify that the subsystem characteristic timescale -- the Josephson frequency -- is to be the shortest one in the entire system, so that observation and control of the system can address only much longer time scales.  Each subsystem then has, in isolation, exactly two conserved quantum numbers: particle number and energy. For a fixed number of particles, the energy may still be varied independently by exciting the Josephson mode; if the rapid Josephson oscillations are averaged over, then states of any energy still count as equilibrium. Thus, the internal Josephson oscillations of each subsystem are to be regarded as heat. 

This identification is bold enough that it may need some defense. The 2-mode Bose-Hubbard model, which has been very extensively studied from many points of view, is certainly a rather minimalist model for a thermal system.  It is technically ergodic, inasmuch as the corresponding classical system explores the entire energy surface (for fixed particle number); but only because that surface is one-dimensional, so that time evolution has no choice but to explore it all. The system is integrable, not chaotic; and chaos is often invoked as a justification for statistical mechanics.  But the precise relevance of chaos to macroscopic behavior, as opposed to time scale separation alone, is far from fully understood. According to the explicit logic normally used to justify ensemble treatments of dynamical systems, and so relate thermodynamics to mechanics, the fact that time averaging reproduces the phase space average over the energy surface should allow us to consider the N-particle excited states of 2-mode Bose-Hubbard as equilibrium states, as far as any measurements or control operations performed on much longer time scales are concerned.

And if the standard appeal to time averaging allows us to claim our N-particle excited Josephson states as equilibria, then it is at least qualitatively reasonable to say that the energy of those excitations is heat. For a macroscopic sample of ordinary gas held in a fixed external potential, with a fixed number of atoms, there is a one-parameter family of equilibrium states, distinguished by total energy (usually expressed as temperature). If we raise this energy without changing the external potential or the particle number, we say that we have added heat to the gas. Indeed this is almost the definition of heat, practically speaking. If heat means anything, it must mean this. 

We therefore argue that the excitations of a two-mode Bose-Hubbard system can indeed provide a simple model for heat, if they are involved in a dynamical process that occurs slowly compared to their Josephson frequency. We arrange for this by placing two such two-mode subsystems in thermal contact, through an additional Josephson tunneling between them, with a much smaller rate coefficient than that of intra-subsystem tunneling.  The two subsystems can thereby exchange particles, on the 'slow' time scale; but they can also exchange heat, as the amplitudes of their fast Josephson oscillations beat slowly back and forth. Our combined four-mode Bose-Hubbard system is thus a candidate minimalist model for heat exchange.

The implications of this model for mesoscopic and quantum thermodynamics depend simply on whether our Josephson heat really behaves as heat.  If it does not, because our integrable system is in some way too trivial, then this will be a useful insight into the precise role of chaos. Even a reductio ad absurdum is still a reductio. But in fact we will show that the beating Josephson excitations do behave very similarly to heat in a recognized, though unusual, form: second sound, which is normally described as a temperature oscillation.  We identify a nonlinear collective effect that makes Josephson oscillations of heat -- 'second Josephson oscillations' -- the slowest collective excitation of the 4-mode system about an excited equilibrium state. Our analytical derivation of this effect, from a simple quantum mechanical Hamiltonian, indicates how thermodynamics may emerge from quantum dynamics in small systems.

\section{The four-mode System}
Our complete system is thus a rather symmetric four-mode Bose-Hubbard model, with two sharply differing tunneling rates.  Such systems may be realized to a good approximation in current cold atom laboratories \cite{Gati07,Bloc05}, either using four closely neighboring microtrap potentials, or with a double well trap holding atoms with two internal states coupled by a Rabi laser \cite{Myat97, Matt99}. Our system can probably also be realized at least classically with superconducting Josephson junctions, but with the cold atom context in mind, we will refer to the particles as atoms.

\begin{figure}[htbp]
\begin{center}
\includegraphics[width=.7\linewidth]{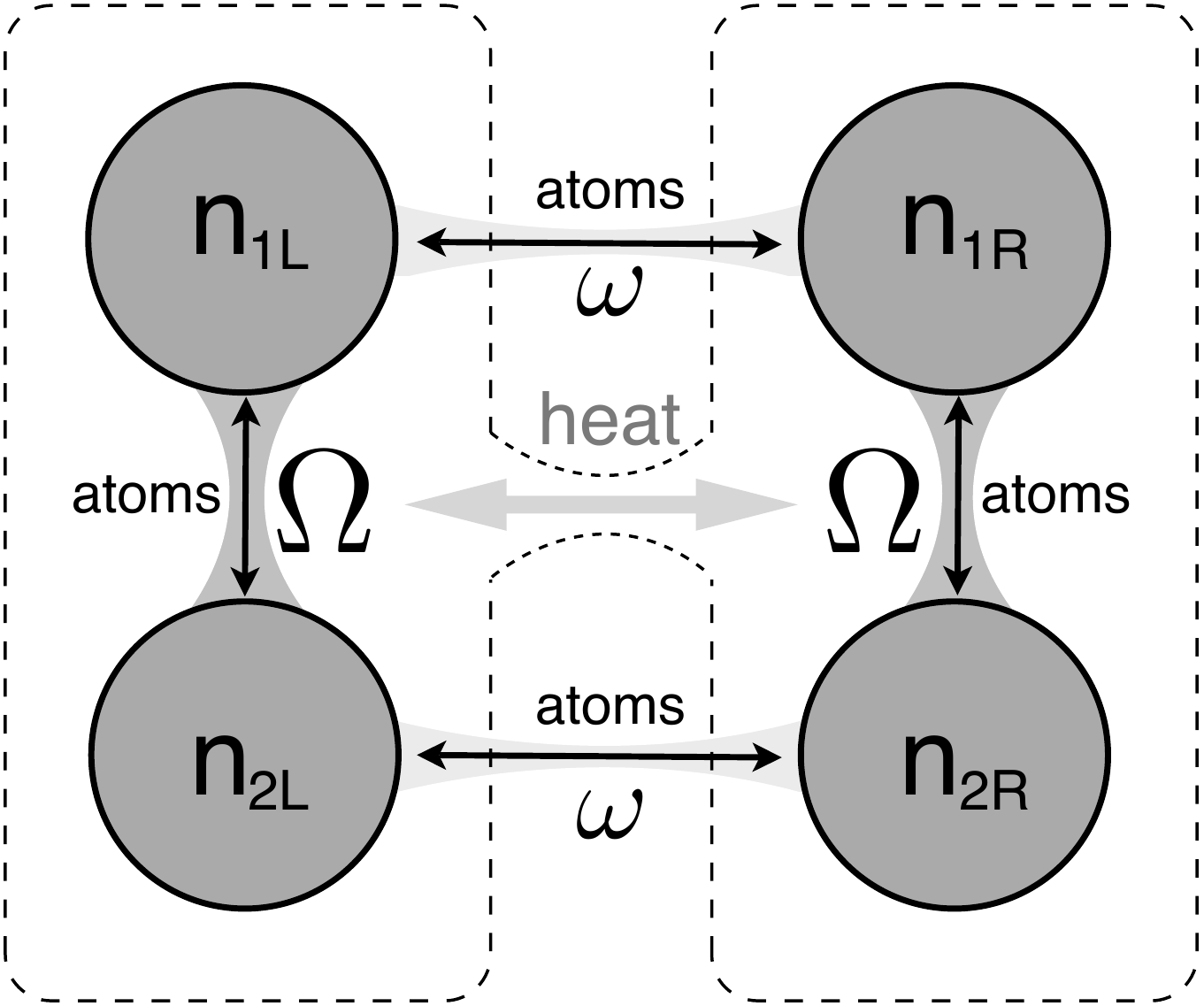}
\caption{Bose-Hubbard system as model for thermal contact.}
\label{default}
\end{center}
\end{figure}

By exploiting the small ratio between the slow and fast tunneling rates, we are able to analyze collective excitations in this model beyond the linear Bogoliubov approximation, and show that the low-frequency behavior of the joint system consists of two Josephson oscillations: the elementary Bogoliubov excitation whereby the two subsystems exchange atoms with each other, and a nonlinear beat mode by which they exchange heat.  Insofar as a Josephson oscillation represents a two-mode analog of Bogoliubov zero sound in a dilute Bose gas, we identify our second Josephson mode as an analog of second sound (see for example \cite{Nozi90}). We offer it as a simple example of a typical phenomenon of mesoscopic thermodynamics.

Our system's four-mode Bose-Hubbard (BH) Hamiltonian with onsite interaction $U$ and index $\alpha=L,R$ is
\begin{eqnarray}\label{Hamiltonian}
 \hat H &=& \hat{H}_{L}+\hat{H}_{R} -\frac{\omega}{2}\left(\hat a_{1L}^{\dagger}\hat a_{1R} + \hat a_{2L}^{\dagger}\hat a_{2R} + \textrm{H. c.}\right)\nonumber\\
 \hat{H}_{\alpha} &=&  -\frac{\Omega}{2}\left(\hat a_{1\alpha}^{\dagger}\hat a_{2\alpha} +\hat a_{2\alpha}^{\dagger}\hat a_{1\alpha} \right) + U\sum_{i=1}^{2} \hat a_{i\alpha}^{\dagger 2}\hat a_{i\alpha}^{2}
\end{eqnarray}
where $\Omega$ is the high tunneling rate within each subsystem pair, and $\omega$ the low rate between the two subsystems.

$\hat{H}$ conserves the total atom number $\hat{N}=\sum_{i, \alpha}\hat{a}_{i\alpha}^{\dagger}\hat{a}_{i\alpha}$, and in the present paper we assume for simplicity an eigenstate of $\hat{N}$ with large eigenvalue $N$.  The global phase zero mode corresponding to $\hat{N}$ is therefore eliminated, and we are left with the three nontrivial tunneling modes illustrated in Fig.~2. 
\begin{figure}[htp]
 \begin{center}
  \includegraphics[scale=0.4]{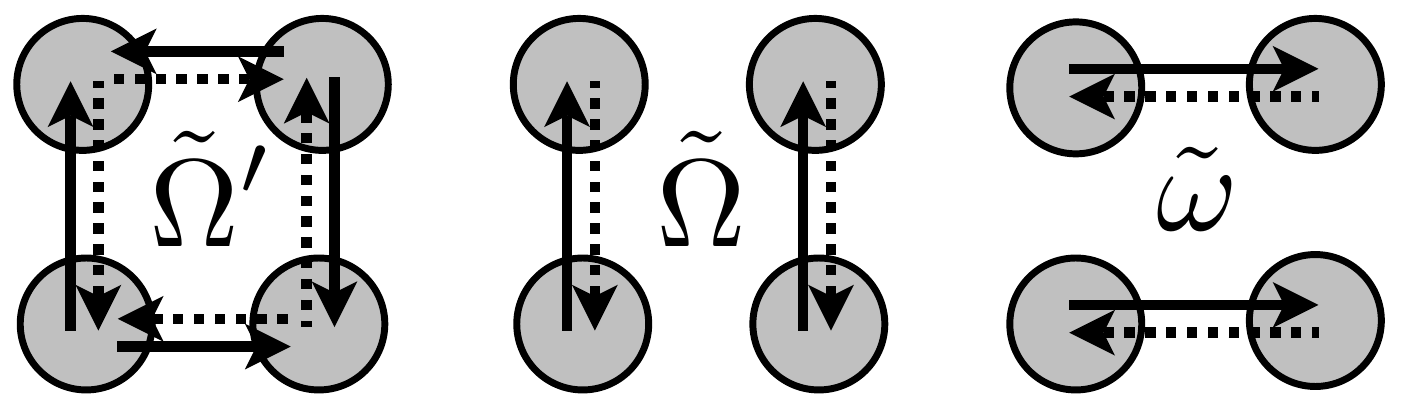}
 \end{center}
 \caption{The three linear Bogoliubov modes.}
\end{figure}
In the linear Bogoliubov approximation the frequencies of these elementary excitations above the $N$-atom ground state are 
\begin{eqnarray}\label{bogfreq}
 \tilde\omega &=& \sqrt{\omega (\omega + UN)},\quad \tilde\Omega = \sqrt{\Omega (\Omega + UN)}\textrm{ and} \nonumber\\
\tilde\Omega' &=& \sqrt{(\Omega+\omega)(\Omega + \omega + UN)}.
\end{eqnarray}
These three modes are Josephson oscillations, and $\tilde\omega$, $\tilde\Omega$, $\tilde\Omega'$ are the three Josephson frequencies for the three independent Josephson oscillations shown in Fig.~2.  As is well known, these collective mode frequencies are shifted away from the pure tunneling (or for internal transitions, Rabi) frequencies $\omega$, $\Omega$ by the nonlinear interactions.  Repulsive interactions ($U>0$) raise the Josephson frequencies, while attractive interactions ($U<0$) lower them.

In the $\omega\ll\Omega$ case we consider, only one of these three elementary modes is slow, and the others are fast. Since excitations of the Bogoliubov modes provide a complete basis for the $N$-particle many-body Hilbert space, it may seem puzzling that there are in fact (as we shall see) two distinct collective modes in this system with low frequencies (of order $\omega$ rather than $\Omega$). This situation is actually common in nonlinear quantum many-body systems, however. The `type II' excitations in the one-dimensional Lieb-Lieniger gas \cite{Lieb63a,Lieb63b}, second sound in superfluid helium \cite{Nozi90}, and even ordinary sound in air, are all examples of low-frequency collective excitations that are not to be found among a system's low-frequency elementary excitations, despite the fact that these are a complete set. 

The resolution of this paradox is that the frequency of beats between our system's two high frequency modes, 
\begin{eqnarray}\label{omegaJ}
	\omega_{J}\equiv (\omega/\tilde\Omega)(\Omega+UN/2)=\tilde\Omega'-\tilde\Omega+\mathcal{O}(\omega^{2}/\Omega)\;,
\end{eqnarray}
is of order $\omega$, and it is the interference beat between the two nearly degenerate high-frequency modes that provides the extra low-frequency collective mode.  What we now show, by calculating to leading order in the small frequency ratio $\omega/\Omega$, but nonlinearly in excitation amplitude, is that within the low-frequency regime this beat mode is an independent excitation in its own right, whose frequency is shifted away from $\omega_{J}$ by nonlinear effects.  

\section{Beyond Bogoliubov}
To go accurately beyond the linear approximation within the standard symmetry-breaking Bogoliubov theory requires careful treatment of zero modes, but it is straightforward if we follow a number-conserving approach \cite{Gard97,Cast98}. Applying the Holstein-Primakoff transformation (HPT) \cite{Hols49}, we re-write $\hat{H}_{\alpha=L,R}$ in terms of the total atom number on each side $\hat{N}_{\alpha}=\sum_{i=1,2}\hat{a}_{i\alpha}^{\dagger}\hat{a}_{i\alpha}$, and atom-moving operators $\hat{a}_{\alpha}^{\dagger},\hat{a}_{\alpha}$ that commute with $\hat{N}_{\alpha}$:
\begin{eqnarray}\label{schwing}
	2\sqrt{\hat{N}_{\alpha}-\hat{a}^{\dagger}_{\alpha}\hat{a}_{\alpha}}\ \hat{a}_{\alpha} \equiv (\hat{a}_{1\alpha}^{\dagger}+\hat{a}_{2\alpha}^{\dagger})(\hat{a}_{1\alpha}-\hat{a}_{2\alpha})\;.
\end{eqnarray}
It is easy to show that $[\hat{a}_{\alpha},\hat{a}^{\dagger}_{\alpha}]=1$ and $[\hat{a}_{\alpha},\hat{N}_{\alpha}]=0$.
In these terms the Hamiltonians of the subsystems read
\begin{eqnarray}
 \hat{H}_{\alpha} &=&  -\frac{\Omega}{2}\hat{N}_\alpha + \Omega \hat{a}_{1\alpha}^{\dagger}\hat{a}_{2\alpha} +\mathcal{O}(U\hat{N}_{\alpha}^{-1})\\
 &+& \frac{U}{2}\hat{N}_\alpha(\hat{N}_\alpha-2) + \frac{U}{2}\hat{N}_\alpha(\hat{a}^\dagger_\alpha+\hat{a}_\alpha)^2 \nonumber\\
 &-& \frac{U}{4}\{ \hat a_\alpha^\dagger + \hat a_\alpha , \hat a_\alpha^\dagger \hat a_\alpha\hat a_\alpha + \hat a_\alpha^\dagger \hat a_\alpha^\dagger \hat a_\alpha \}\nonumber.
\end{eqnarray}

We can then implement the Bogoliubov formalism by defining $\hat{a}_\alpha = u_{\alpha}\hat{b}_\alpha + v_{\alpha}\hat{b}_\alpha^\dagger$ and choosing $u_{\alpha},v_{\alpha}$ to diagonalize the quadratic terms in $\hat{H}_{\alpha}$ while maintaining $[\hat{b}_{\alpha},\hat{b}^{\dagger}_{\alpha}]=1$.  

Since we are interested in dynamics slow compared to $\Omega$, and in Josephson oscillations of finite but not extreme amplitude, we will apply the rotating wave approximation (RWA) by dropping from $\hat{H}_{\alpha}$ all terms that do not commute with the leading term proportional to $\hat{b}^{\dagger}_{\alpha}\hat{b}_{\alpha}$.  Assuming the regime of large occupations for simplicity, we omit corrections of order $U\hat{N}_{\alpha}^{-1}$, and obtain for $\hat{H}_{L,R}$
\begin{eqnarray}\label{Halpha1}
\hat{H}_{\alpha} &\to& -\frac{\Omega}{2}\hat{N}_{\alpha}+\frac{U}{2}\hat{N}_{\alpha}(\hat{N}_{\alpha}-2) \\
&+&\!\!\!\!\sqrt{\Omega(\Omega+2U\hat{N}_{\alpha})}\ \hat{b}^{\dagger}_{\alpha}\hat{b}_{\alpha} -\frac{U}{4}\frac{4\Omega +2U\hat{N}_{\alpha}}{\Omega + 2U \hat{N}_{\alpha}}\hat{b}^{\dagger2}_{\alpha}\hat{b}^{2}_{\alpha}\;.\nonumber
\end{eqnarray}

In the quadratic term of (\ref{Halpha1}) we note the familiar Josephson frequency of the excitations created by $\hat{b}^{\dagger}_{\alpha}$, which we have been unable to resist naming `josons'.  The co-efficient of the term quartic in $\hat{b}_{\alpha}$ is in general of the same order as $U$, but opposite sign; this reflects the well-known fact that the period of nonlinear Josephson oscillations lengthens with amplitude \cite{Smer97}.  In quasiparticle terms, we may say that where atoms repel each other, josons attract; and \textit{vice versa}.

To simplify the description of the slow tunneling between the $L,R$ subsystems, we now perform a similar HPT for our two atom numbers, while assuming that our system is in an eigenstate of the conserved total $\hat{N}_{L}+\hat{N}_{R}$, with eigenvalue $N$.  This defines a number-conserving operator $\hat{a}$, which satisfies $[\hat{a},\hat{a}^{\dagger}]=1$; it transfers atoms between the $L$ and $R$ subsystems.  In the large $N$ limit that we assume for simplicity, this provides
\begin{eqnarray}
 \hat{N}_{L,R} &=& \frac{1}{2}[ N \pm N^{1/2}(\hat a^\dagger + \hat a)] 
 \end{eqnarray}
up to terms of $\mathcal{O}(\hat{a}^{\dagger}\hat{a}^{2}N^{-1/2})$.  We can therefore re-write our full Hamiltonian (\ref{Hamiltonian}) in terms of $\hat{b}_{L,R}$, $\hat{a}$, and $N$.  
 
Since $\tilde\Omega$ is on the order of the problem's high frequency $\Omega$, we see that to leading (zeroth) order in 
$\omega/\Omega$, our entire Hamiltonian is simply $\tilde\Omega\hat{J}$, for
\begin{eqnarray}\label{J}
	\hat{J}\equiv\hat{b}^{\dagger}_{L}\hat{b}_{L}+\hat{b}^{\dagger}_{R}\hat{b}_{R}.
\end{eqnarray}
In classical terms, the `total joson number' $\hat{J}$ is an adiabatic invariant.  By means of an RWA we can now compute low frequency dynamics nonlinearly in excitation amplitude, but to leading order in small frequency ratios, by dropping all terms in $\hat{H}$ that do not commute with $\hat{J}$.  Since $\hat{b}^{\dagger}_{\alpha}\hat{b}_{\alpha}$ are not separately adiabatic invariant, we must retain terms such as $\hat{b}^{\dagger}_{L}\hat{b}_{R}$.  Our result is
\begin{eqnarray}\label{Hfull}
 \hat H &\to& \omega\hat a^\dagger\hat a + \frac{U}{4} N(\hat a^\dagger + \hat a)^2\nonumber \\
 &&- \frac{\omega_J}{2} \left(\hat b_L^\dagger \hat b_R + \hat b_R^\dagger \hat b_L\right) + \frac{U_J}{2}\sum_{\alpha=L,R} \hat b_\alpha^\dagger \hat b_\alpha^\dagger \hat b_\alpha\hat b_\alpha \nonumber\\
 &&+ \frac{U}{2}\sqrt{\frac{\Omega N}{\Omega + UN}}(\hat a^\dagger + \hat a)(\hat b_L^\dagger\hat b_L -\hat b_R^\dagger\hat b_R) \nonumber\\
\omega_{J}&\equiv&\omega\frac{\Omega+UN/2}{\sqrt{\Omega (\Omega + UN)}} \nonumber\\
U_{J}&\equiv&-\frac{U}{4}\frac{4\Omega +UN}{\Omega + UN}\;.
\end{eqnarray}
In the first two lines we can recognize two different forms of the standard single Josephson junction Hamiltonian.  The first line provides Josephson oscillations of atoms between $L$ and $R$ subsystems, with Josephson frequency $\tilde\omega = \sqrt{\omega(\omega+UN)}$, while the second implies \textit{Josephson oscillations of josons}.  The two Josephson modes are exactly analogous, except that $U_{J}$ and $U$ are of opposite sign, as anticipated above following (5).  The third line couples the atom and joson modes, and arises because the frequency of fast Josephson oscillations in each $L,R$ subsystem depends on the respective atom number.

\section{Second Josephson Oscillations}

We can now perform a final HPT to express $\hat{b}_{\alpha}$ in terms of $\hat{J}$ and a $J$-conserving operator $\hat{b}$ that transfers josons between subsystems.  As an adiabatic invariant, $\hat{J}$ is just as good a conserved quantity within the low-frequency regime as $\hat{N}$ itself, and so we may let $\hat{J}\to J$ in the low-frequency theory.  To identify collective excitations, we then linearize (\ref{Hfull}) in $\hat{a}, \hat{b}$ to obtain for fixed $N,J$
\begin{eqnarray}\label{Hfinal}
 \hat{H}_{\textrm{lin}} &=& \omega\hat a^\dagger\hat a + \frac{U}{4} N(\hat a^\dagger + \hat a)^2\nonumber \\
 &+& \omega_J \hat b^\dagger\hat b + \frac{U_J}{4} J (\hat b^\dagger + \hat b)^2\nonumber\\
 &+& \frac{U}{2}\frac{\Omega}{\tilde\Omega}(NJ)^{1/2}(\hat a^\dagger + \hat a)(\hat b^\dagger + \hat b). 
\end{eqnarray}
In terms of the atom Josephson frequency $\tilde\omega=\sqrt{\omega(\omega+UN)}$ and the new Josephson frequency for josons
\begin{equation}
 \tilde{\omega}_{J}=\sqrt{\omega_{J}(\omega_{J}+U_{J}J)},
\end{equation}
the decoupled collective mode frequencies in (\ref{Hfinal}) are
\begin{equation}\label{finalmodes}
 \tilde{\omega}_\pm^{2} = \frac{ \tilde{\omega}^2+ \tilde{\omega}_{J}^2}{2} \pm \left[\left(\frac{ \tilde{\omega}^2 -  \tilde{\omega}_J^2}{2}\right)^{2} +\frac{\omega\omega_J \Omega U^{2}NJ}{\Omega + UN}\right]^{1/2}\!\!\!\!\!\!\!\!.
\end{equation}

We observe that in the low-amplitude regime $J\ll N$, either $UN\gg \omega$ so that $\tilde\omega \gg\tilde{\omega}_{J}$; or else $UN \lesssim \omega$ and thus $\tilde\omega \approx \tilde{\omega}_{J}$ so that the third line in (\ref{Hfinal}) is much smaller than the first two.  Thus for $J\ll N$ the low frequency atom Josephson frequency $\tilde\omega_{+}\doteq\tilde{\omega}$ is always essentially unchanged from the standard Bogoliubov theory indicated in Fig.~(2); but it is joined by a second Josephson oscillation whose frequency $\tilde{\omega}_{-}$ can be significantly lower (for $U>0$), because $U_{J}$ is negative.  This is our main result.  

Note that the linearization that yields (\ref{Hfinal}) is different from the standard Bogoliubov approximation of retaining only quadratic terms in fluctuations of the original $\hat{a}_{i\alpha}$ about the mean-field ground state. It is a resummed Bogoliubov formalism, in which certain dominant nonlinear interactions among atoms have been included in the linearized interactions among josons, which are 
nonlinearly related to atoms through their definition via the HPT.  One may also say that we do pursue the standard (conserving) Bogoliubov approach, but applied to a two-mode system of interacting conserved atoms and josons. In this sense the extra low-frequency mode is a non-elementary collective mode in the full theory, but an elementary excitation within the adiabatic effective theory for low frequency dynamics.

Although our derivation is post-Bogoliubov in the sense of being nonlinear, it is not essentially non-classical, because we have assumed large $N$, and because all our HPT and RWA steps are exactly paralleled in classical adiabatic theory.  We can therefore test our theory of second Josephson oscillations numerically by solving the four-mode Gross-Pitaevskii nonlinear Schr\"odinger equation for the classical limit of our system. We excite intra-subsystem Josephson excitations with an oscillating drive, as follows.
We relax to an $N$-particle mean field ground state, then excite josons by adding a time-dependent potential tilt to each subsystem, which varies periodically with the Josephson frequency $\tilde\Omega$. We excite the second Josephson mode by giving the tilt drives slightly different amplitudes for the $L$ and $R$ pairs, so that slightly more josons are excited on one side than on the other.  The resulting evolution, after the drive is turned off, is shown in Fig.~3 for a typical case.  
\begin{figure}[htb]
 \begin{center}
 \includegraphics[width=1\linewidth]{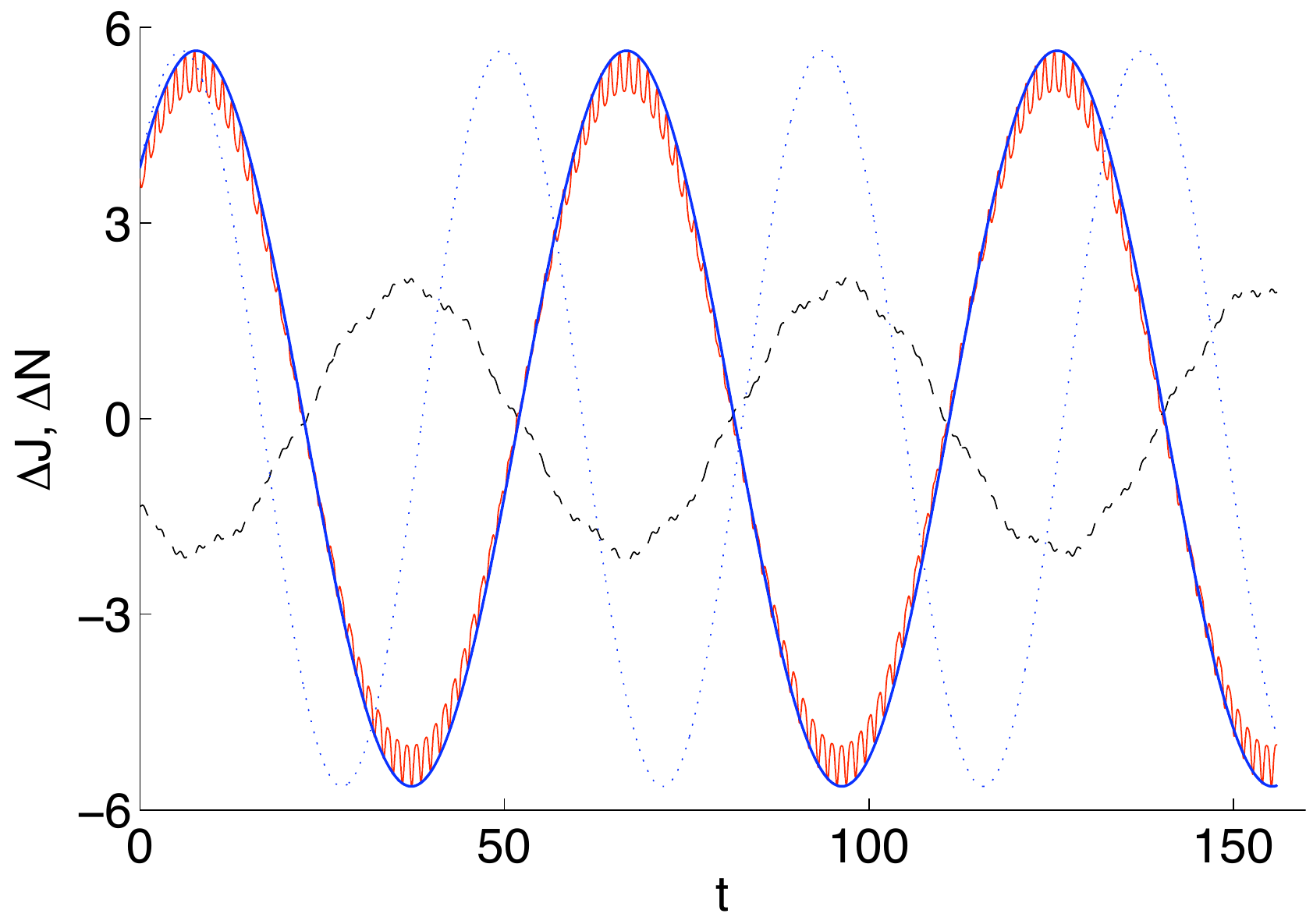}
 \end{center}
 \caption{Time evolution of $\Delta J =  b_L^\dagger b_L - b_R^\dagger b_R$ and $\Delta N$ in the mean-field system for the parameters $\Omega = 1$, $\omega = 0.1$, $UN = 5$, $N=10^{4}$ and $J = 139$.  Red solid line: numerical mean-field $\Delta J$. Blue solid line: oscillation at second Josephson frequency $\tilde{\omega}_-$ computed in the text, fitted for phase.  Blue dotted line: oscillation at beat-frequency $\omega_J$. Black dashed line: numerical $\Delta N$.}
 \label{numerics}
\end{figure}

The agreement with our linearized theory is excellent, and the nonlinear shifting of the second Josephson frequency well below the trivial Bogoliubov beat is clear.  Some atom population oscillation between $L$ and $R$ subsystems occurs at frequency $\tilde\omega_{+}$, this mode having been slightly excited by our driving procedure, but the main atom oscillation is exactly opposite in phase to the joson oscillation.  We note that while in incompressible superfluids second sound includes no include significant particle density change, the coupling of density and heat oscillations is expected for second sound in compressible superfluids \cite{Grif97}.  We interpret the coupling between the slow atom and joson oscillations as a simple example of the interconversion of heat and work.

\section{Discussion}
The question whether our proposed oscillation of heat carries oscillating entropy is as much about the microscopic definition of entropy, as about the validity of our identification of heat.  Our josons do carry at least one physically meaningful form of entropy: the von Neumann entropy of the single-particle reduced density matrices $R_{L,R}$, projected onto either half of our system.  These are simply two-by-two matrices, diagonal in the basis of even and odd modes excited by $\hat a_{\pm \alpha} = (\hat{a}_{1\alpha}\pm\hat{a}_{2\alpha})/\sqrt{2}$:
\begin{eqnarray}\label{SPDM}
R_{\alpha} &=& \frac {\langle \hat a_{\pm\alpha}^\dagger \hat a_{\pm\alpha}\rangle}{\langle \hat N_{\alpha}\rangle}=\left( \begin{array}{cc}1-P_{\alpha} & 0 \\ 0 & P_{\alpha}
\end{array} \right) \textrm{ with}\\
P_{\alpha} &=& \frac{1}{N_{\alpha}}\left[ \frac{\Omega + UN_{\alpha}}{\sqrt{\Omega(\Omega+2UN_{\alpha})}}\left(J_{\alpha}+\frac{1}{2}\right) - \frac{1}{2} \right]\!\!, \label{Palpha}
\end{eqnarray}
for $J_{\alpha} = \langle \hat b_\alpha^{\dagger} \hat b_{\alpha}\rangle$.  The local single-particle entropies $S_{L,R}$ on each side,
\begin{eqnarray}\label{entropy}
	S_{\alpha}=-k_{\textrm B} [P_{\alpha}\ln P_{\alpha}+(1-P_{\alpha})\ln(1-P_{\alpha})]
\end{eqnarray}
are genuine entropies inasmuch as they characterize the first order coherence, or equivalently the condensate depletion, within the two-mode Bose-Hubbard subsystems.  They are direct analogs, for their respective two-mode subsystems, of the single-particle entropy which has been used since Boltzmann to represent thermodynamic entropy in statistical mechanics.  

Since $S_{\alpha}$ is a simple function of $J_{\alpha}$, it is clear that our josons carry entropy at least in this single-particle sense.  One could even consider letting $J_{\alpha}\to\hat{b}^{\dagger}_{\alpha}\hat{b}_{\alpha}$ and $N_{\alpha}\to\hat{N}_{\alpha}$ in (\ref{Palpha}) and (\ref{entropy}), in order to define Hermitian entropy operators $\hat{S}_{\alpha}$ which would be ordinary quantum mechanical observables. One of the basic questions in quantum thermodynamics is whether quantum entropy should be an observable operator just like quantum energy, with state-independent eigenvalues that appear as measurements with state-dependent probabilities, or whether it should remain a fundamentally statistical quantity like the full von Neumann entropy $S=-k_{\textrm B}\mathrm{Tr}\hat{\rho}\ln\hat{\rho}$, which is a nonlinear function of the density operator $\hat{\rho}$ that represents the quantum state of an open system.  Our simple model thus addresses this question by suggesting a form of entropy operator, whose performance in the role of thermodynamic entropy may be assessed experimentally.

The general implication of our work for mesoscopic quantum thermodynamics would seem to be that it is rather easy to identify First Law phenomenology, whereby parametric oscillations among high-frequency internal degrees of freedom provide those extra contributions to the low-frequency energy budget which in thermodynamics are called heat.  Inasmuch as this heat oscillates in our case, rather than diffusing, we must recognize that Second Law phenomenology, with irreversible spontaneous processes, may be more elusive in small quantum systems.  On the other hand, the full range of quantum dynamics in even the four-mode Bose-Hubbard model, in parameter ranges not considered here, is rich beyond the scope of this paper.  It includes quantum chaos \cite{Stoe99}, whose role in mesoscopic irreversibility amply deserves further study.

\section{Acknowledgment}

Support from the Deutsche Forschungsgemeinschaft via the Graduiertenkolleg ``Nichtlineare Optik
und Ultrakurzzeitphysik'' is gratefully acknowledged.


\begin{thebibliography}{10}

\bibitem{Deut91}
J.~M. Deutsch
  Phys. Rev. A {\bf 43}  (1991) 2046

\bibitem{Sred94}
M. Srednicki
  Phys. Rev. E {\bf 50}  (1994) 888

\bibitem{Kino06}
T. Kinoshita, T. Wenger and D.~S. Weiss
  Nature {\bf 440}  (2006) 900

\bibitem{Rigo08}
M. Rigol, V. Dunjko and M. Olshanii
  Nature {\bf 452}  (2008) 854

\bibitem{Ecks09}
M. Eckstein, M. Kollar and P. Werner
  Phys. Rev. Lett. {\bf 103}  (2009) 056403

\bibitem{Rigo09a}
M. Rigol
  Phys. Rev. Lett. {\bf 103}  (2009) 100403



\bibitem{Land80}
L.~D. Landau and E.~M. Lifshitz, 
  \textit{Statistical Physics, Part 1}  (Pergamon Press, Oxford, 1980), Chap. 1, \textsection1

\bibitem{Bloc05}
I. Bloch,
  Nature Physics {\bf 1}  (2005) 23

\bibitem{Gati07}
R. Gati and M.~K. Oberthaler,
  J. Phys. B  {\bf 40}  (2007)   R61

\bibitem{Myat97}
C.~J. Myatt, E.~A. Burt, R.~W. Ghrist, E.~A. Cornell and C.~E. Wieman,
  Phys. Rev. Lett. {\bf 78}  (1997)   586

\bibitem{Matt99}
M.~R. Matthews, B.~P. Anderson, P.~C. Haljan, D.~S. Hall, M.~J. Holland, J.~E. Williams, C.~E. Wieman and E.~A. Cornell,
  Phys. Rev. Lett. {\bf 83}  (1999)   3358

\bibitem{Nozi90}
P. Nozi\`{e}res and D. Pines, 
  \textit{The Theory of Quantum Liquids} (Addison-Wesley, Redwood City, 1990), Vol. II.

\bibitem{Lieb63a}
E.~H. Lieb and W. Liniger,
  Phys. Rev. {\bf 130}  (1963)   1605

\bibitem{Lieb63b}
E.~H. Lieb,
  Phys. Rev. {\bf 130}  (1963)  1616

\bibitem{Gard97}
C.~W. Gardiner,
  Phys. Rev. A  {\bf 56}  (1997)   1414

\bibitem{Cast98}
Y. Castin and R. Dum,
  Phys. Rev. A  {\bf 57}  (1998)   3008

\bibitem{Hols49}
T. Holstein and H. Primakoff,
  Phys. Rev. {\bf 58}  (1949)   1098

\bibitem{Milb97}
G.~J. Milburn, J. Corney, E.~M. Wright and D.~F. Walls,
  Phys. Rev. A  {\bf 55}  (1997)   4318

\bibitem{Smer97}
A. Smerzi, S. Fantoni, S. Giovanazzi and S.~R. Shenoy,
  Phys. Rev. Lett. {\bf 79}  (1997)   4950

\bibitem{Grif97}
A. Griffin and E. Zaremba,
  Phys. Rev. A  {\bf 56}  (1997)   4839

\bibitem{Stoe99}
H.~J. St\"{o}ckmann,
\textit{Quantum Chaos} (Cambridge University Press, Cambridge, 1999)

\end{thebibliography}
\end{document}